\documentclass[a4paper,11pt]{article}
\usepackage{amsmath,amsthm,amssymb,slashed}
\usepackage{pos}
\usepackage{subfig}
\usepackage{mathrsfs}

\usepackage[symbol]{footmisc}
\renewcommand{\thefootnote}{\fnsymbol{footnote}}

\newcommand{\be}{\begin{equation}}
\newcommand{\ee}{\end{equation}}
\newcommand{\bea}{\begin{eqnarray}}
\newcommand{\eea}{\end{eqnarray}}
\newcommand{\ba}{\begin{array}}
\newcommand{\ea}{\end{array}}

\title{Non-unitarity and sterile neutrinos at the DUNE near detector}

\author*[a]{Salvador Urrea\footnote[2]{Work done in collaboration with: Pilar Coloma, Jacobo López-Pavón and Salvador Rosauro-Alcaraz}}

\affiliation[a]{Instituto de Física Corpuscular, Universidad de Valencia \& CSIC, 
Edificio Institutos de Investigaci\'on, Calle Catedr\'atico Jos\'e Beltr\'an 2, 46980 Paterna, Spain}

\emailAdd{salvador.urrea@ific.uv.es}

\abstract{We study the capabilities of the DUNE near detector to probe the 3+1 sterile
formalism and the non-unitarity of the leptonic sector. We add to the current analyses in the literature, the use of
the charged current events for the $\nu_{\tau}$ appearance channel and the consideration
of the energy spectral uncertainty (shape uncertainty) for
the different channels which plays an important role, especially at the near
detector, and has been usually overlooked in the literature. We find that even with
this more conservative and realistic approach, we still obtain an improvement in
the sensitivity with respect to the current bounds.}

\FullConference{%
  *** The European Physical Society Conference on High Energy Physics (EPS-HEP2021), ***\\
  *** 26-30 July 2021 ***\\
  *** Online conference, jointly organized by Universität Hamburg and the research center DESY ***
}

\begin{document}
\maketitle

\section{Introduction}

The new generation of long-baseline neutrino oscillation experiments such as DUNE are expected to probe the robustness of the three neutrino family oscillation picture to a high level of accuracy. They will also give stronger bounds and/or possibly detect non-standard deviations from the standard picture.

The ability to probe new physics in oscillations is going to be heavily affected by systematic uncertainties coming from our lack of knowledge, both theoretical and experimental, of the cross sections and fluxes involved. In the case of the near detector (ND), the lack of another detector which would reduce uncertainties, results in larger systematics. We will see that the uncertainties in the energy spectra will play the main role in these searches.

\section{Theoretical framework and notation}
\label{sec:notation}

In this section, we will briefly introduce the two theoretical scenarios that we will consider in our sensitivity analysis: non-unitarity and sterile neutrinos. For a more detailed discussion see \cite{Coloma_2021}.
 
A simple way to account for light neutrino masses is to add singlet fermions to the SM field content. The matrix mixing between the mass and flavor basis can be expressed as:

\be
\label{eq:U}
\mathcal{U} = \begin{pmatrix}
N & \Theta  \\
R & S \\
\end{pmatrix} \, ,
\ee
Here $N$ is a $3\times3$ non-unitary matrix corresponding to the PMNS active-light sub-block, and $\Theta$ is the $3\times n$ sub-block parameterizing the mixing between active and heavy neutrinos, with $n$ being the number of new states. The $R$ ($S$) sub-block gives the mixing between the sterile and light (heavy) states and does not play any role in neutrino oscillations.

The value of the mass of the new states will determine whether they can be produced in the neutrino beam or not, changing the phenomenology we expect to observe. We have two distinct cases: if their mass lies above the production threshold of the neutrino source they cannot be produced, this scenario is usually referred to as Non-unitarity (NU) and typically mass scales above the electroweak scale are assumed; whereas if the new states are kinematically accessible, they can be produced in the neutrino beam leading to new oscillation frequencies driven by the new masses. We refer to the latter as light sterile neutrinos.

\subsection{Parameterization}

We will parametrize the deviations from unitarity of the matrix $N$ as follows:  
\be
N  = (I-T) U 
\label{eq:N}
\ee
where $T$ is given by:
\be
T = \begin{pmatrix}
\alpha_{ee} & 0 & 0\\
\alpha_{\mu e} & \alpha_{\mu \mu} & 0\\
\alpha_{\tau e} & \alpha_{\tau \mu} & \alpha_{\tau \tau}
\end{pmatrix},
\label{eq:alpha}
\ee
and $U$ plays the role of the standard(unitary) PMNS matrix up to small corrections encoded in $\alpha_{\gamma\beta}$.

\subsection{Non-Unitarity from new physics above the electroweak scale}
\label{sec:notation-NU}

In this scenario, the heavy states are integrated out from the low energy spectrum and are thus not kinematically accessible in the experiment. At very short distances, as the ones in the DUNE ND, the standard oscillation do not have time to develop, yielding the following simple expressions for the oscillations probabilities:
\bea
\label{eq:PNUL0}
P_{\mu e}(L=0) &=& |\alpha_{\mu e}|^2,\nonumber\\
P_{\mu\tau}(L=0) &=& |\alpha_{\tau\mu}|^2.
\eea

\subsection{Sterile neutrinos \& Non-Unitarity from new physics at low scales}
\label{sec:notation-sterile}

As we mentioned, sterile neutrinos are kinematically accesible and we expect a new oscillation frequency. For simplicity, we consider the $3+1$ scenario in which only one new neutrino is introduced.

At very short baselines, we find the following simple expressions for the oscillation probabilities:
\bea
\label{eq:sterile-eff-angles}
P_{\gamma\beta} =& 
 \sin^22\vartheta_{\gamma\beta}\sin^2\left(\dfrac{\Delta m^2_{41}L}{4E}\right), \quad & \rm{with} \;\; \sin^22\vartheta_{\gamma\beta} \equiv 4\,|\mathcal{U}_{\beta 4}|^2|\mathcal{U}_{\gamma 4}|^2 , \nonumber\\
P_{\beta\beta} = & 1 - \sin^22\vartheta_{\beta\beta}\sin^2\left(\dfrac{\Delta m^2_{41}L}{4E}\right), \quad & \rm{with} \;\; \sin^2\vartheta_{\beta\beta} \equiv |\mathcal{U}_{\beta 4}|^2 \, .
\eea 
In the averaged-out regime ($\Delta m^2_{41} L/E \gg 1$) the $\sin^2\left(\dfrac{\Delta m^2_{41}L}{4E}\right)$ average out to 1/2 leading to following energy independent expressions:
\bea
\label{eq:PsterilesL0_1}
P_{\mu e}= 2\,|\mathcal{U}_{\mu 4}|^2|\mathcal{U}_{e 4}|^2 = 2|\alpha_{\mu e}|^2, P_{\mu\tau}&=& 2\,|\mathcal{U}_{\mu 4}|^2|\mathcal{U}_{\tau 4}|^2 = 2|\alpha_{\tau\mu}|^2
\eea 

Notice that, barring the factor 2, the new physics effects here are identical to those given in Eq. \ref{eq:PNUL0}. Thus we can consider this regime as a low scale source of non-unitarity

\section{Results}
\label{sec:results}
In this section, we will present our main results of our paper, for the rest of the results including the sensitivity analysis for NSI in detection and production check \cite{Coloma_2021}.

Before going to the results, we will comment on what type of systematics will be important to have sensitivity to the different scenarios we have just described. At the DUNE ND the sensitivity will be dominated by the spectral information, since even for a value of $\alpha_{\gamma\beta}$ that saturates the present bound the signal is much smaller than the background. The sensitivity comes mainly from the differences in energy shape between the background and signal. Shape uncertainties are generally overlooked in the literature.

\subsection{Non-unitarity and sterile averaged-out regime }

\begin{figure}[ht!]
\begin{center}
  \includegraphics[width=1\textwidth]{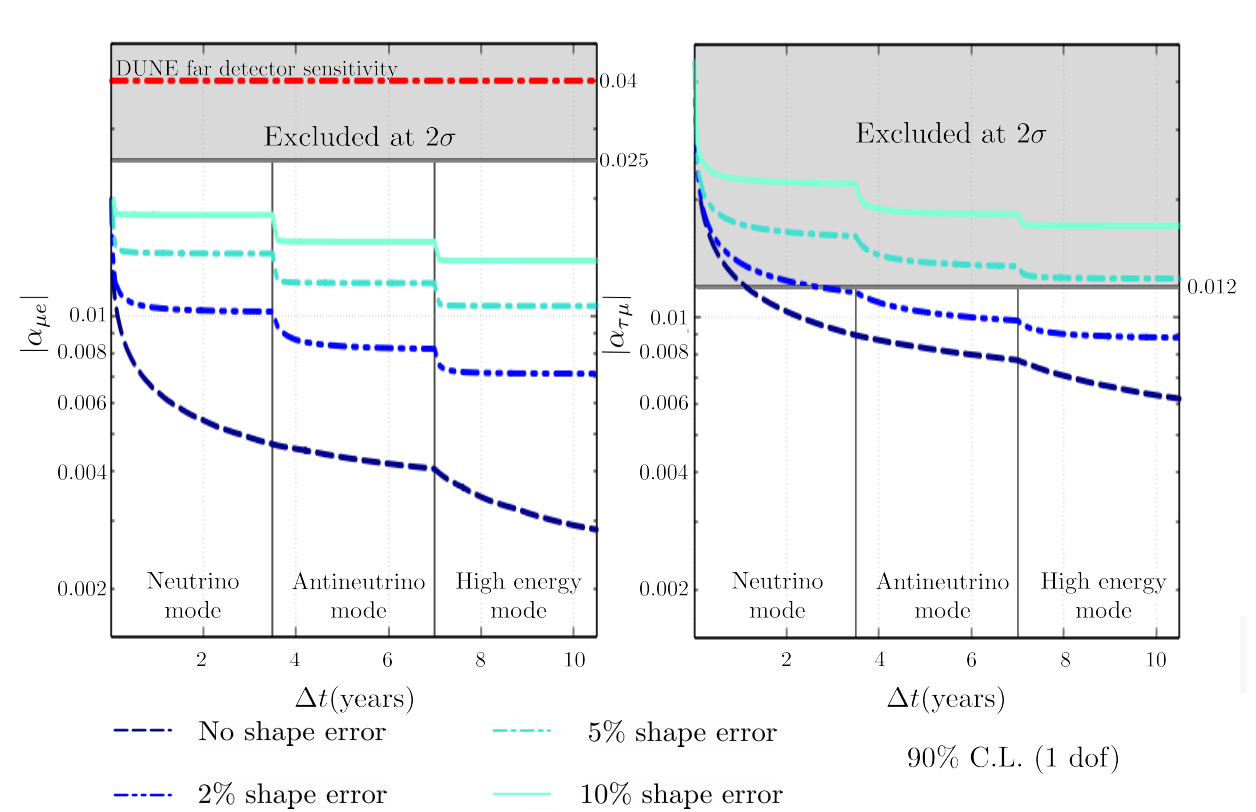} 
\end{center}
\caption{\label{fig:NU} Sensitivity to the off-diagonal NU parameters  $\alpha_{\mu e}$ (left panel) and $\alpha_{\tau\mu}$ (right panel). The lines show the sensitivity at 90\% CL for 1 degree of freedom (d.o.f.) as a function of the running time $\Delta t$, for different choices of the systematic uncertainties. The vertical lines indicate the changes between neutrino and antineutrino running modes (in the nominal beam scenario) as well as the change to the high energy beam. For reference, the shaded gray areas show the region disfavored at 90\% CL by current constraints. For $\alpha_{\mu e}$, the dot-dashed red line indicates the expected sensitivity using the far detector data (taken from Ref.~\cite{Blennow:2016jkn}), while for $\alpha_{\tau\mu}$ the expected sensitivity would be worse than the range shown in the figure and is therefore not shown.}
\end{figure}

Comparing Eq.\eqref{eq:PNUL0} with Eq.\eqref{eq:PsterilesL0_1}, we see that in the appearance channels there is only a factor 2 difference, and therefore, the results can be rescaled from one scenario to the other. Even though, in oscillations they only differ by this factor 2, the bounds that apply for both scenarios are different. In the Non-unitarity scenario we have very strong bounds from high-precision measurements of electroweak processes which are not expected to be improved by near future oscillation experiments\cite{Blennow:2016jkn}.
However, these constraints do not apply when the sterile neutrinos are kinematically accessible. Indeed, the bounds we show in Fig \ref{fig:NU}, which come from oscillation experiments are considerably weaker and apply to the average-out regime of the steriles. See \cite{Coloma_2021} for details.  

Therefore, our results shown in Fig.~\ref{fig:NU} for $\alpha_{\mu e}$ (left panel) and $\alpha_{\tau\mu}$ (right panel), respectively, correspond to the averaged-out regime of the steriles. The different lines show the results for different choices of systematic uncertainties as indicated by the labels, as a function of running time. 

\subsection{Sterile neutrino oscillations}

In the sterile neutrino scenario, the DUNE ND will be sensitive to different oscillation channels. Each of them will be sensitive to a different set of mixing matrix elements as described in Sec.~\ref{sec:notation}, see Eq.~\eqref{eq:sterile-eff-angles}.
 
\begin{figure}[ht!]
\begin{center}
  \includegraphics[width=0.65\textwidth]{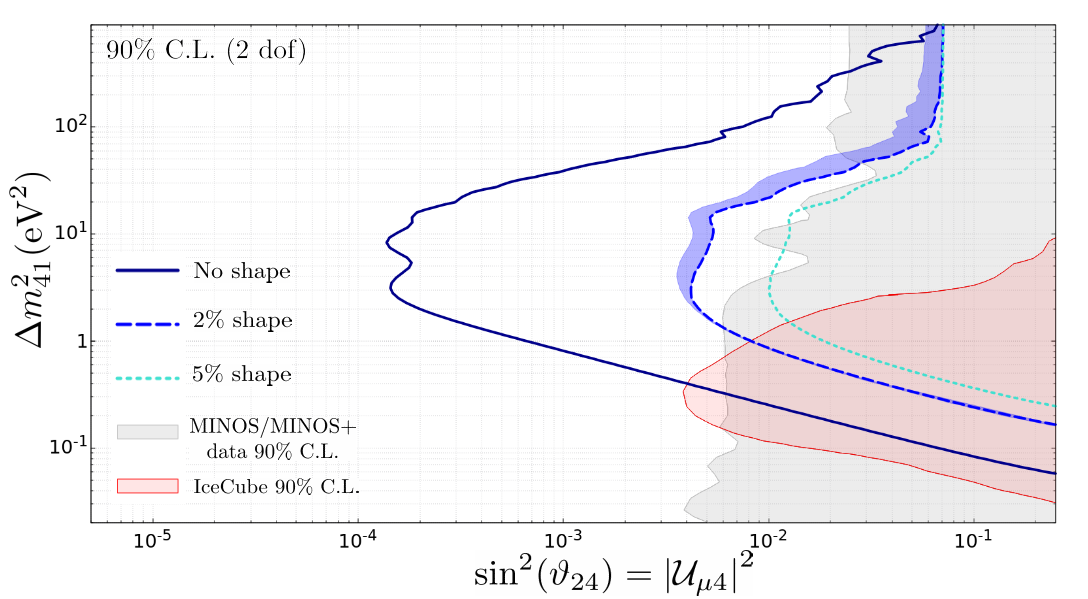}
  \includegraphics[width=0.65\textwidth]{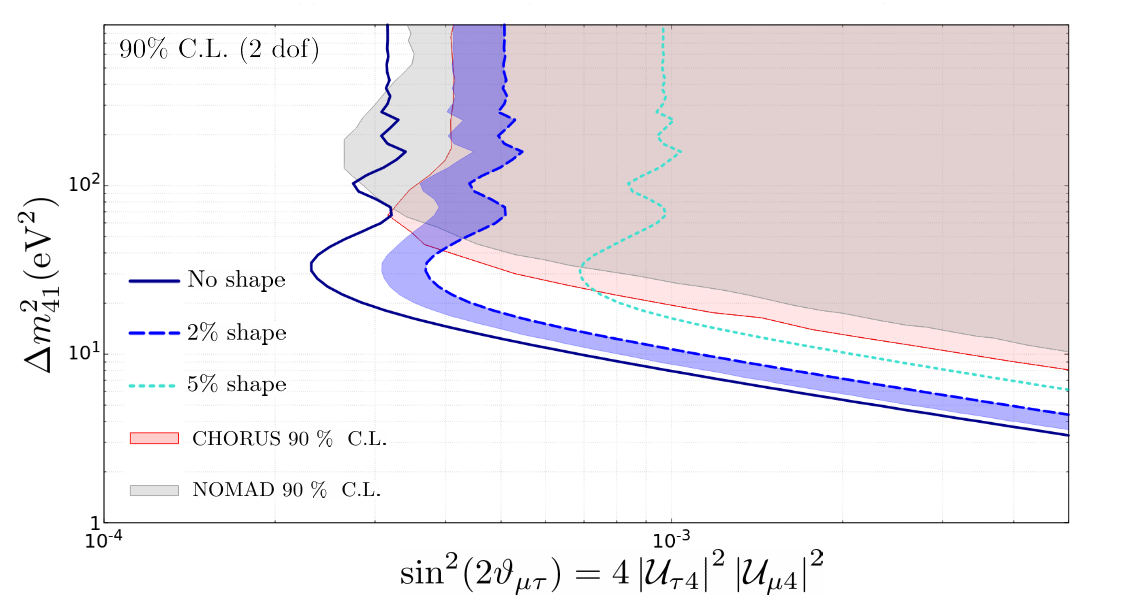}
\end{center}
\caption{\label{fig:sterile} Expected sensitivity to the sterile neutrino scenario, for oscillations in the $P_{\mu\mu}$ and $P_{\mu\tau}$  channels for the top and lower panels, respectively. The shaded regions show current constraints from other experiments, while the colored lines indicate the sensitivity for the DUNE ND, for different choices of systematic uncertainties as indicated by the legend. Finally, the colored blue bands indicate the increase in sensitivity due to the addition of 3.5 years of data taken in the high energy mode.  }
\end{figure}

In Fig.~\ref{fig:sterile} we present the results obtained for the sensitivity for $\nu_\mu$ dissapearance and $\nu_\tau$ appearance channels. Shape uncertainties have a large impact on the results in all channels. It is worth mentioning that the most dramatic effect occurs for the sensitivity via $P_{\mu\mu}$, where shape uncertainties at the level of 2\% lead to a decrease in sensitivity of over an order of magnitude with respect to the scenario where only normalization uncertainties are considered. This is expected, since the $\nu_{\mu}$ flux constitutes the largest component of the incoming flux and the same relative error results in a larger absolute error.

\begin{figure}[ht!]
\begin{center}
  \includegraphics[width=1\textwidth]{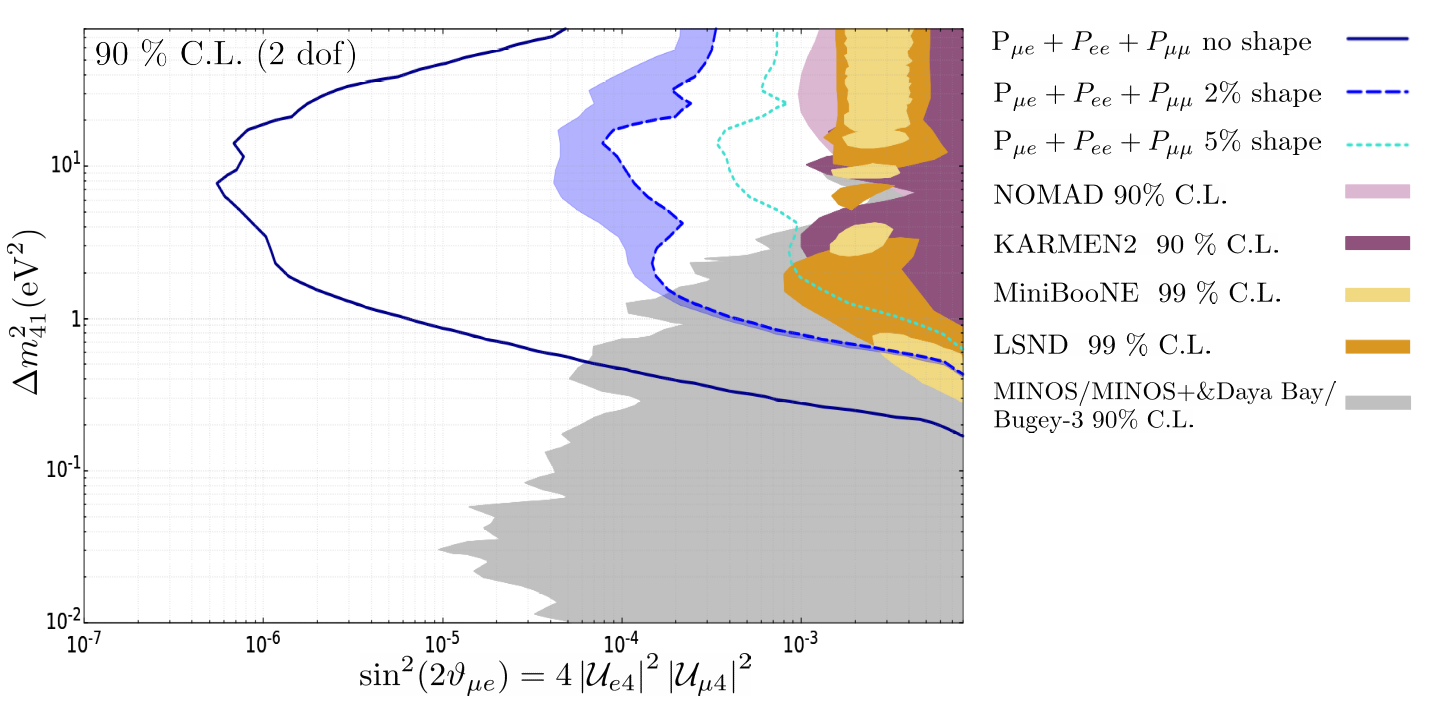}
\end{center}
\caption{\label{fig:sterile-Pmue} Expected sensitivity to the sterile neutrino scenario, for oscillations in the $P_{\mu e}$ channel. The colored lines indicate the sensitivity for the DUNE ND, for different choices of systematic uncertainties. In each case, the region to the right of the lines would be disfavored at 90\% CL (2 d.o.f.). Finally, the shaded band to the left of the dashed lines indicate the increase in sensitivity due to the addition of  3.5 years of data taken in the HE mode.}
\end{figure}

Figure~\ref{fig:sterile-Pmue}, on the other hand, shows the DUNE ND sensitivity to combination of mixing matrix elements that appear in $P_{\mu e}$. Looking at Eq. \ref{eq:sterile-eff-angles} we can see that the same parameters are accessible by the combination of $P_{e e}$ and $P_{\mu \mu}$. Therefore, the three channels together will give us the best sensitivity. From these results, we see that for DUNE ND is expected to probe the regions favoured by the LSDN and MiniBooNE results.

 \section{Conclusions}

In this work we have studied the sensitivity to new physics affecting neutrino oscillations in the DUNE ND. We have presented a conservative but realistic approach, including shape uncertainties, for different scenarios and channels. Even though, the expected sensitivity is, therefore, reduced we generally expect an improvement over present bounds. With this work we want to stress the importance of the reduction of these type of systematics from an experimental and also theoretical point of view.

\textit{Acknowledgments}: I acknowledge support
from Generalitat Valenciana through the plan GenT program (CIDEGENT/2018/019) and
from the Spanish MINECO under Grant FPA2017-85985-P.

{\footnotesize
\bibliographystyle{JHEP}
\bibliography{references}}

\end{document}